\documentclass[prl,aps,twocolumn,amsmath,amssymb,superscriptaddress,intlimits]{revtex4-1}
\usepackage{bm,latexsym,mathrsfs,enumerate,color}
\usepackage[mathcal]{euscript}
\usepackage{graphicx}
\usepackage{adjustbox}
\usepackage{array,ragged2e}
\usepackage[breaklinks=true,unicode=true,urlcolor = blue,colorlinks = true,citecolor = blue,linkcolor = blue]{hyperref}

\renewcommand{\vec}[1]{\bm{#1}}
\bibpunct{[}{]}{,}{n}{}{}

\newcommand{\s}{\alpha}
\renewcommand{\v}{v}
\renewcommand{\c}{\xi}

\begin{document}

\title{
Solitary wave excitations of skyrmion strings in chiral magnets}

\author{Volodymyr P. Kravchuk}
\affiliation{Leibniz-Institut f{\"u}r Festk{\"o}rper- und Werkstoffforschung, IFW Dresden, D-01171 Dresden, Germany}
\affiliation{Bogolyubov Institute for Theoretical Physics of National Academy of Sciences of Ukraine, 03680 Kyiv, Ukraine}

\author{Ulrich~K.~R{\"o}{\ss}ler}
\affiliation{Leibniz-Institut f{\"u}r Festk{\"o}rper- und Werkstoffforschung, IFW Dresden, D-01171 Dresden, Germany}

\author{Jeroen~van~den~Brink}
\affiliation{Leibniz-Institut f{\"u}r Festk{\"o}rper- und Werkstoffforschung, IFW Dresden, D-01171 Dresden, Germany}
\affiliation{Department of Physics, Washington University, St. Louis, MO 63130, USA}

\author{Markus Garst}
\affiliation{Institut f\"ur Theoretische Physik, Technische Universit\"at Dresden, 01062 Dresden, Germany}

\date{\today}

\begin{abstract}
Chiral magnets possess topological line excitations   
where the magnetization within each cross section forms a skyrmion texture. We study analytically and numerically the low-energy, non-linear dynamics of such a skyrmion string in a field-polarized cubic chiral magnet, and we demonstrate that it supports solitary waves. Theses waves are in general non-reciprocal, i.e., their properties depend on the sign of their velocity $v$, but this non-reciprocity diminishes with decreasing $|v|$. An effective field-theoretical description of the solitary waves is derived that is valid in the limit $v \to 0$ and gives access to their profiles and their existence regime.  Our analytical results are quantitatively confirmed with micromagnetic simulations for parameters appropriate for the chiral magnet FeGe. Similarities with solitary waves found in vortex filaments of fluids are pointed out. 
\end{abstract}


\maketitle

Manifestations of one-dimensional topological objects can be found in diverse systems such as domain walls in magnetic films, vortex filaments in classical as well as quantum fluids, or cosmic strings in the universe \cite{Malozemoff79,Pismen99}. Usually, these string-like objects can elastically bend and twist, and this elasticity also determines their dynamical behavior at low energies. This results in collective low-amplitude excitations, for example, 
Kelvin waves \cite{Thomson1880} that propagate along vortex filaments and play a key role in the decay of superfluid turbulence \cite{Svistunov95}. Certain filaments can also support non-linear solitary waves, which are spatially confined excitations that maintain a constant amplitude. For vortex filaments in fluids, solitary waves were theoretically predicted by Hasimoto \cite{Hasimoto72} and subsequently observed experimentally \cite{Hopfinger1982}. 

Topological strings of a different kind arise in cubic chiral magnets like MnSi, FeGe or Cu$_2$OSeO$_3$. Here, the competition between exchange and Dzyaloshinskii-Moriya interactions stabilises magnetic skyrmions, i.e., two-dimensional topological textures of the magnetization \cite{Bogdanov1989}. These skyrmions extend along the third direction forming topological strings. Skyrmion strings align with the applied magnetic field $\vec H$, and, under certain conditions, they condense into a hexagonal lattice forming a thermodynamically stable phase \cite{Muehlbauer09,Seki:2012ie,Seidel_Bauer2016}. 

The non-trivial topology of two-dimensional skyrmion textures has direct consequences for their magnetization dynamics. It is reflected by their  Thiele equation of motion that predicts an efficient coupling to spin currents with interesting spintronic applications \cite{Thiele73,Jonietz:2010fy}, for recent reviews see \cite{Nagaosa13,Wiesendanger16,Fert17,JIANG20171,Everschor-Sitte18}. It also leads to an emergent electrodynamics for electrons as well as spin waves, that scatter off skyrmions, resulting in topological Hall effects \cite{Neubauer09,Lee09a,Schulz12,Iwasaki2014,Mochizuki:2014fj}. In addition, skyrmions possess characteristic internal degrees of freedom like a breathing mode, that can be studied with magnetic resonance spectroscopy
\cite{Mochizuki:2012kc,Onose2012,Schwarze2015,Lin:2014bv,Schuette14,Kravchuk18,Garst:2017}.

\begin{figure}
\includegraphics[width=0.8\linewidth]{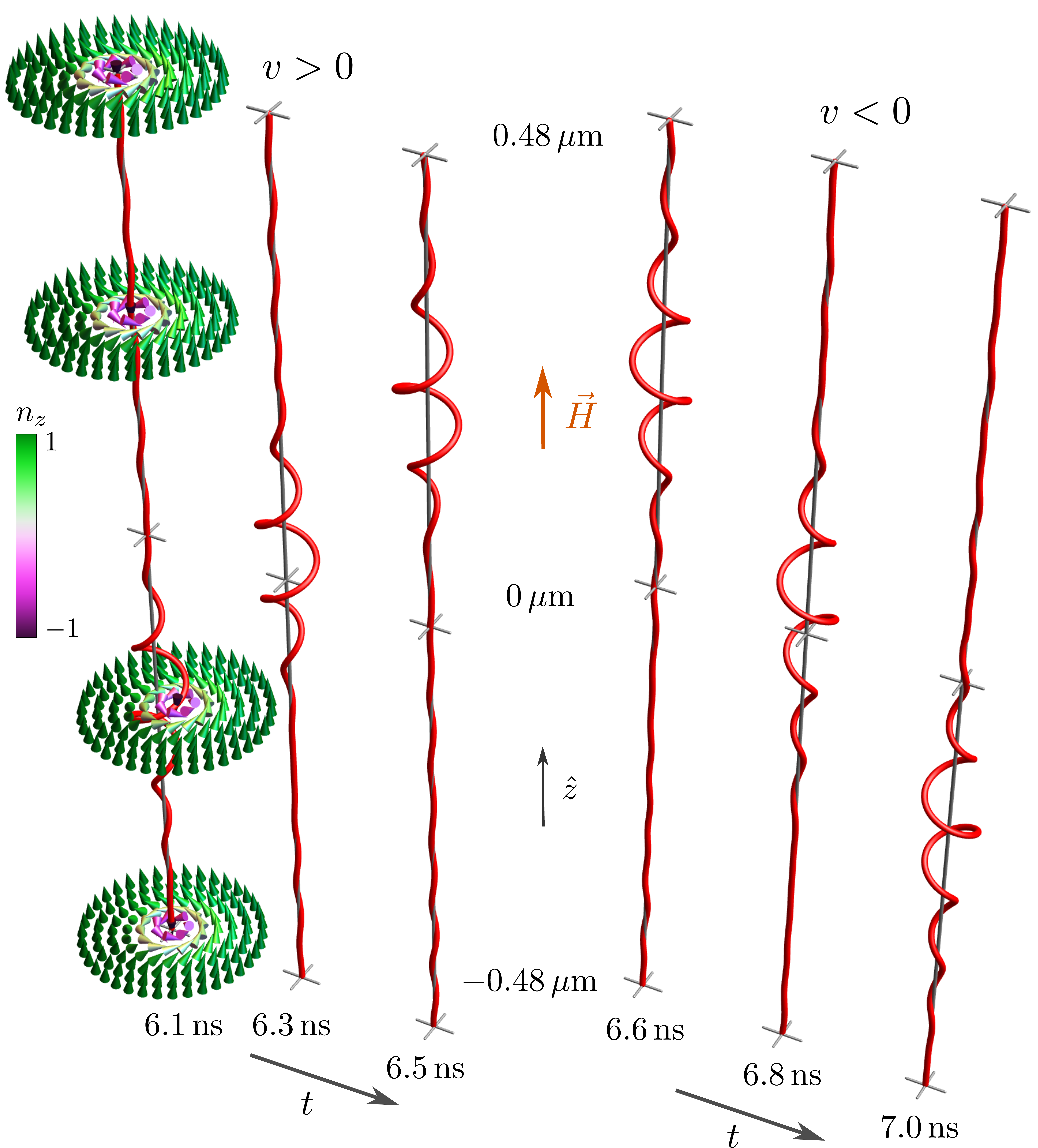}
\caption{
Solitary wave excitation of an isolated skyrmion string, that is aligned with the magnetic field $\vec H = \hat{\vec z} H$, propagating in a direction parallel ($v>0$) and antiparallel ($v<0$) to $\vec H$. The figure is produced by micromagnetic simulation with parameters $\omega_{c2}/(2\pi) = 10.4$ GHz and $2\pi/Q = 70$ nm typical for the chiral magnet FeGe \cite{Haraldson78,Nagaosa13}. The dimensionless field is $h = 2.16$ corresponding to $\mu_0 H \approx 0.8$ T for $g \approx 2$ resulting in a skyrmion string radius of approximately~$5$ nm. Solitary waves with amplitude $R_0 \approx 5.8$ nm are created at time $t=0$ and propagate with velocity $|v| \approx 1$ km/s. Periodic boundary conditions are employed, and the red line is defined according to Eq.~\eqref{Position}. Lengths in the $(x,y)$-plane are upscaled by a factor of five for visualization.
	}
\label{fig:1}
\end{figure}

Even richer is the dynamics of three-dimensional skyrmion strings due to their additional degrees of freedom.
Recently, it was demonstrated that spin waves generally propagate along these strings in a non-reciprocal fashion \cite{Xing19,Lin2019,Seki19}. The elastic response of skyrmion strings gives rise to a non-reciprocal, non-linear Hall effect \cite{Muehlbauer_2016,Yokouchi18}.  Moreover, the merger of strings is necessarily accompanied with singular Bloch points that play the role of magnetic monopoles in the emergent electrodynamics \cite{Milde13,Schuette14b,Lin16}. 
In chiral magnets polarized either by a magnetic field or a uniaxial anisotropy, single skyrmion strings exist as topologically stable line excitations. Focussing on cubic chiral magnets, we demonstrate in the present work that solitary waves can propagate along such isolated strings, see Fig.~\ref{fig:1},  efficiently transmitting energy, linear and angular momentum.

When damping is neglected, the magnetization dynamics is described by the Landau-Lifshitz equation $\partial_t \vec n = - \gamma (\vec n \times \vec B_{\rm eff})$ for the orientation of the magnetization represented by the unit vector $\vec n$ where $\gamma=g\mu_B/\hbar$ is the gyromagnetic ratio. The effective magnetic field, $\vec B_{\rm eff} = - \frac{1}{M_s} \frac{\delta V}{\delta \vec n}$, derives from the potential $V = \int d\vec r\, \mathcal{V}$; for cubic chiral magnets the potential density is given by 
\begin{align} \label{Theory}
\mathcal{V}(\vec n, \nabla \vec n) = A (\partial_i \vec n)^2 + D \vec n (\nabla \times \vec n) - \mu_0 M_s  H n_z,
\end{align}
with the exchange stiffness $A$, the Dzyaloshinskii-Moriya interaction $D$, the magnetic constant $\mu_0$, the saturation magnetization $M_s$, and the magnetic field $\vec H = \hat{\vec{z}} H$ that defines the $z$-axis. Dipolar interactions are neglected for simplicity. The scales for frequency and wavevectors are given by $\omega_{c2} = \gamma D^2 /(2 A M_s)$ and $Q = D/(2A)$, respectively, and the theory \eqref{Theory} only depends on a single dimensionless parameter $h = \gamma \mu_0 H/\omega_{c2}$ parameterizing the strength of the magnetic field.

For $h > 1$, the ground state of the theory \eqref{Theory} is field-polarized, $n_z = 1$, and we focus on this parameter regime in the following. This field-polarized state possesses a static skyrmion string excitation aligned with the field $\vec H$ \cite{Bogdanov94}. This is a smooth magnetic texture with a quantized topological charge $\mathcal{N}_{\rm top}=\int dx dy\, \rho_{\rm top} = - 1$ for all values of $z$, i.e., for each cross section perpendicular to $\vec H$. Here, $\rho_{\rm top}(\vec r) = \frac{1}{4\pi} \vec n (\partial_x \vec n \times \partial_y \vec n)$ is the topological charge density within the $(x,y)$-plane. 

\begin{figure}
\includegraphics[width=0.9\linewidth]{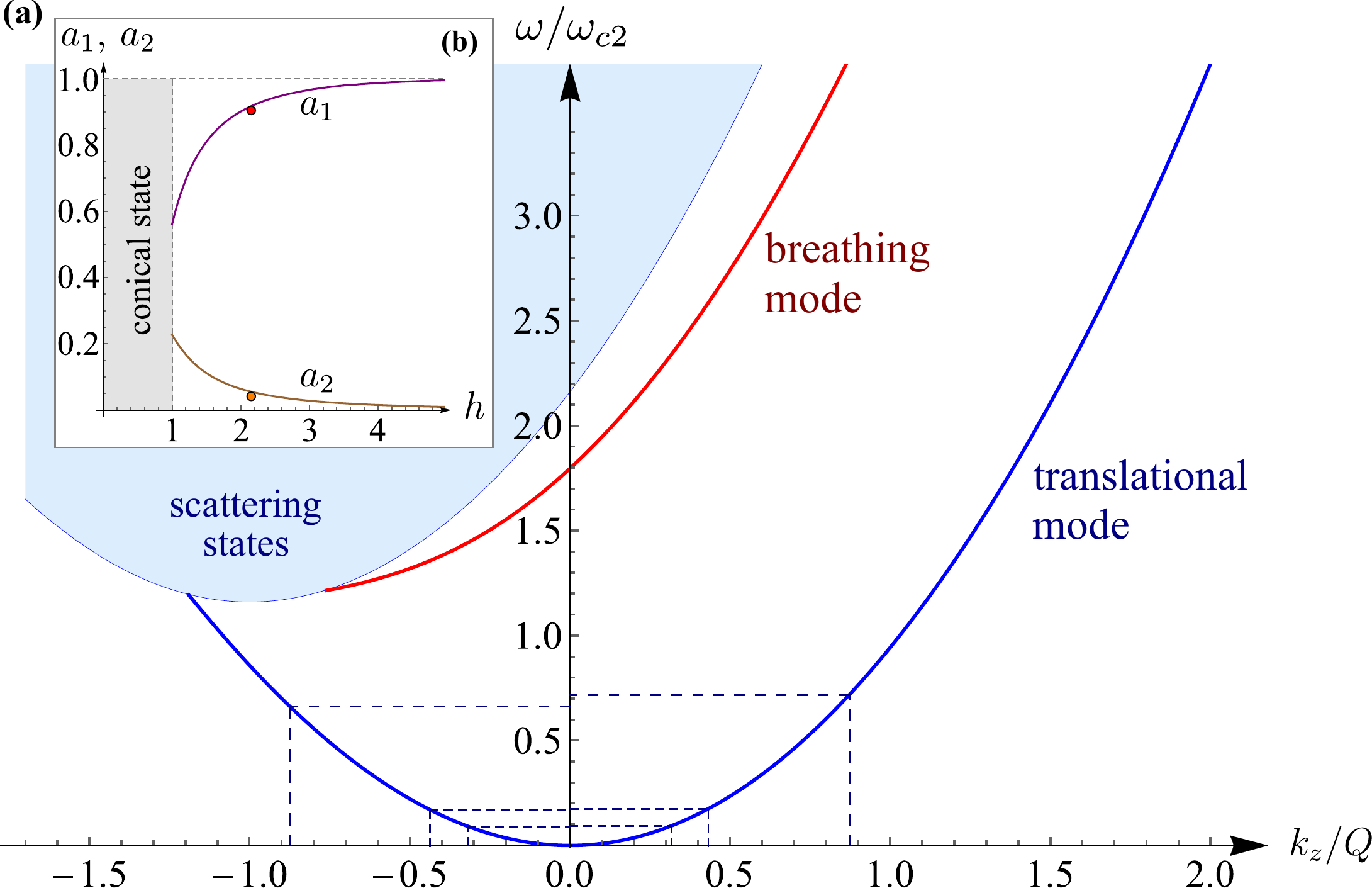}
\caption{(a) Linear spin wave spectrum of a skyrmion string for wavevectors $k_z$ along the string and an applied magnetic field $h=2.16$. The spectrum of the low-energy translational mode has the form $\omega(k_z)/\omega_{c2} \approx  a_1 (k_z/Q)^2 + a_2 (k_z/Q)^3 + ...$ for small $k_z$. The non-reciprocity due to $a_2$ is small for wavevectors $|k_z| \ll k^*$ where $k^* = a_1 Q/a_2$ ($k^*/Q=22.02$ for $h=2.16$). (b) Field dependence of the parameters $a_1$ and $a_2$. The dashed lines in (a) as well as the points marked in (b) refer to 
Figs.~\ref{fig:3} and \ref{fig:4}.
}
\label{fig:2}
\end{figure}

We first discuss the dynamical excitations of the skyrmion string on the level of linear spin-wave theory. The excitation spectrum of the skyrmion was studied analytically in Refs.~\onlinecite{Schuette14,Kravchuk18}. Extending this analysis to the skyrmion string, one obtains the spectrum as a function of wavevector $k_z$ along the field \cite{Lin2019}, see Fig.~\ref{fig:2} and the supplementary materials \cite{SI} for details. The spectrum lacks mirror symmetry with respect to $k_z$, i.e., it is non-reciprocal due to the Dzyaloshinskii-Moriya interaction. The shaded region contains the scattering states, that are gapped for any $h > 1$. In addition, there are magnon-skyrmion bound states with wave functions localized to the skyrmion string. For the range of $k_z$ shown in Fig.~\ref{fig:2}, there exist two dispersive bound states: the breathing mode, that possesses a finite energy at $k_z = 0$, and the translational Goldstone mode. The spectrum of the latter is gapless, $\omega(k_z) \propto k_z^2$ for $|k_z| \ll k^*$,  due to the translational invariance of the theory \eqref{Theory}  \cite{Kobayashi14a}; for the definition of $k^*$ see Fig.~\ref{fig:2}.

The degree of freedom associated with the Goldstone mode is the linear momentum of the skyrmion texture within the $(x,y)$-plane. This linear momentum corresponds to a collective coordinate $\vec{R}=X\hat{\vec x}+Y\hat{\vec y}$, and
it is given by the first-moment of the topological charge density 
\cite{Papanicolaou91,Schuette14,Kravchuk18}
\begin{align} \label{Position}
\vec R(z,t) =  
\frac{1}{\mathcal{N}_{\rm top}} \int dx dy\, \left(x\hat{\vec x}+y\hat{\vec y} \right)\, \rho_{\rm top}(\vec r,t).
\end{align}
We demand that  the skyrmion string does not bend back so that its linear momentum is uniquely defined for each value of $z$. The dynamics of  $\vec R$ is governed by the Thiele equation \cite{Thiele73} that reads in the absence damping $\vec G \times \partial_t \vec R = \vec F$. The gyrocoupling vector $\vec G = \hat{\vec z} 4\pi \mathcal{N}_{\rm top} M_s/\gamma$ is proportional to the topological charge $\mathcal{N}_{\rm top}$, and $\vec F$ is a force that acts on the string. 

In the absence of external forces, $\vec F = - \frac{\delta V_{\rm eff}}{\delta \vec R}$ is generated by the skyrmion string itself and derives from an effective potential $V_{\rm eff} = \int dz \mathcal{V}_{\rm eff}$ with a local density $\mathcal{V}_{\rm eff}$, that can be phenomenologically constructed using symmetry considerations. Due to translational invariance $\mathcal{V}_{\rm eff}$ cannot depend on $\vec R$ itself but only on its derivatives with respect to $z$, i.e., $\partial_z\vec R$, $\partial^2_z\vec R$, etc. We will limit ourselves to the case of small $|\partial^2_z\vec R|/|\partial_z\vec R| \ll k^*$ so that second and higher order derivatives can be omitted. Moreover, due to  invariance of the theory \eqref{Theory} with respect to combined rotations of real and spin space around the $z$-axis the effective potential is only a function of $(\partial_z \vec R)^2$. Assuming analyticity, we can approximate the potential for small $(\partial_z \vec R)^2$ by its Taylor expansion, and we obtain $\mathcal{V}_{\rm eff} \approx \frac{1}{2} K_1 (\partial_z \vec R)^2 - \frac{1}{4} K_2 (\partial_z \vec R)^4$.

\begin{figure}
\includegraphics[width=0.8\linewidth]{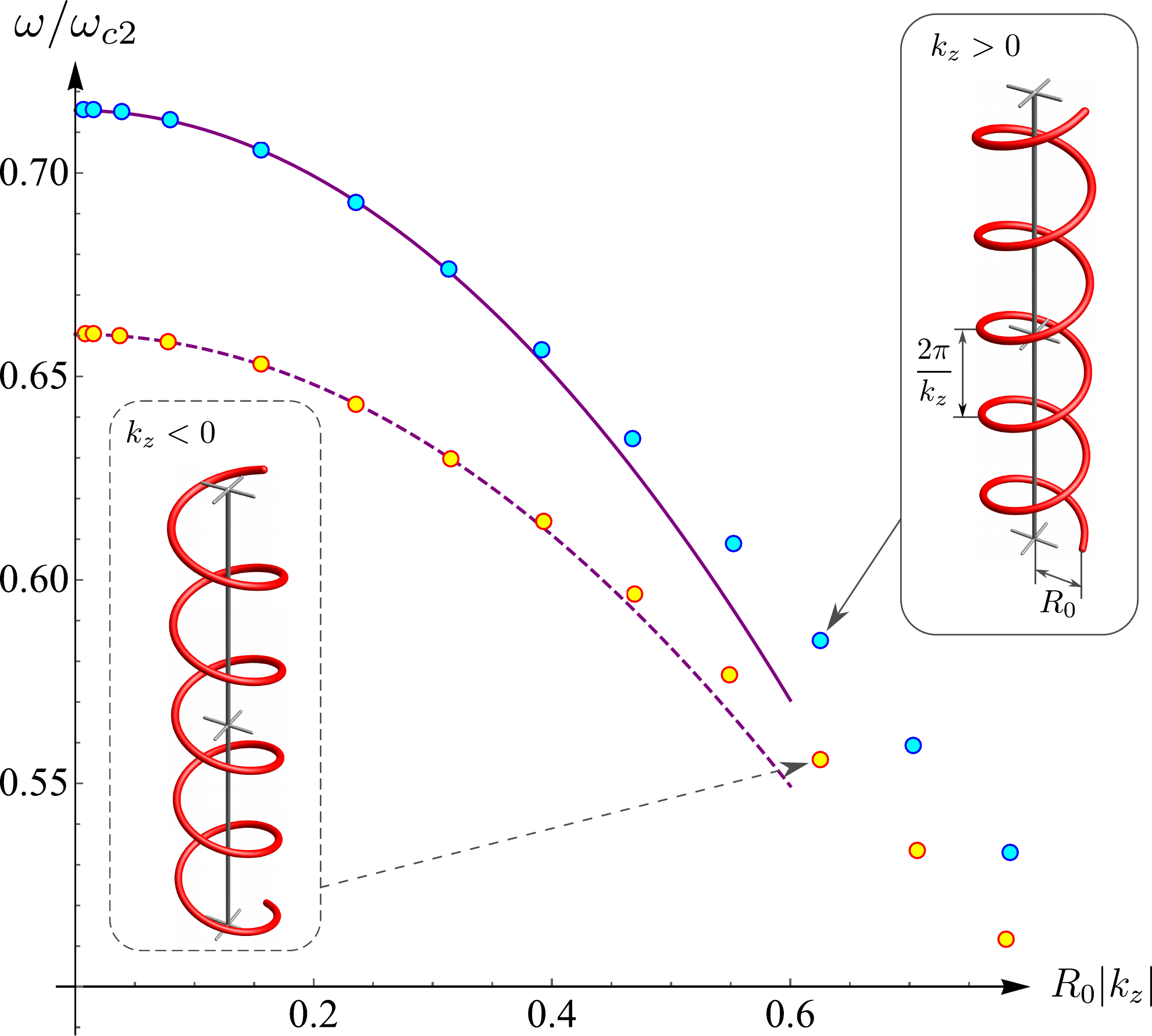}
\caption{Non-linear spin waves of the skyrmion string studied with micromagnetic simulations for $h=2.16$. The energies of low-energy waves, $X + \mathrm{i} Y = R_0 e^{\mathrm{i} k_z z - \mathrm{i} \omega t}$, with fixed wavevector $k_z=0.87 Q$ but for various amplitudes $R_0$ between 0.1 nm and 10 nm are fitted to $\omega/\omega_{c2} = a_1 (k_z/Q)^2 + a_2 (k_z/Q)^3 - (k_zR_0)^2 \left[b_1(k_z/Q)^2 + b_2 (k_z/Q)^3\right]$, and we obtain $a_1\approx0.903, a_2\approx0.041, b_1\approx0.467, b_2\approx0.071$. The values for $a_1$ and $a_2$ are in good agreement with linear spin wave theory, see dots in the inset of Fig.~\ref{fig:2}(b). The coefficients $a_2$ and $b_2$ account for the leading non-reciprocal corrections, that are small for $|k_z| \ll k^*$ and are omitted in Eq.~\eqref{Lagrange}. The amplitude for both simulation snapshots shown in the insets is $R_0 = 8$ nm where lengths in the $(x,y)$-plane are upscaled by a factor of five.
}
\label{fig:3}
\end{figure}

The elastic coefficient $K_1$ quantifies the stiffness of the string, and it is positive in order to guarantee stability.  Importantly, the leading-order non-linearity possesses a coefficient $K_2 > 0$. In the framework of non-linear spin wave theory, this non-linearity arises from  fluctuations of the stiffness $K_1$ that, in second-order perturbation theory, gives rise to a positive $K_2$. Here, we determine both $K_1 = a_1 |\vec G| \omega_{c2}/Q^2$ and $K_2 = b_1 |\vec G| \omega_{c2}/Q^2$ numerically with the help of micromagnetic simulations of non-linear spin waves, see Fig.~\ref{fig:3}. For a magnetic field $h = 2.16$ we obtain $a_1 \approx 0.903$ and $b_1 \approx 0.467$.

In the following, it is convenient to introduce the dimensionless wave function 
\begin{align} \label{WF}
\psi = \sqrt{\frac{2 b_1}{a_1}} Q \left(X + \mathrm{i}\, Y\right).
\end{align}
Using dimensionless time, $x_0 = 2 a_1 t \omega_{c2}$, and length, $x_1 = z Q$, 
the Thiele equation reduces to the Euler-Lagrange equation for the  Lagrange density
\begin{align} \label{Lagrange}
\mathcal{L} = \frac{i}{2}\Big(\psi^* \partial_{0} \psi -  \psi \partial_{0} \psi^* \Big) - \frac{1}{2} |\partial_{1}\psi|^2 + \frac{1}{8} |\partial_{1}\psi|^4.
\end{align}
We arrive at the result that the low-energy dynamics of the skyrmion string is described by a non-linear Schr\"odinger equation 
with an attractive interaction $|\partial_{1}\psi|^4$ involving the derivative of the wavefunction.

Now, we demonstrate that the theory \eqref{Lagrange} possesses solitary waves, and its elementary conservation laws are sufficient to derive their properties. The particle density $j_{0} = |\psi|^2$ obeys the conservation law $\partial_\mu j_\mu = 0$ with the associated current $j_{1}$ along the $z$-axis. The density $|\psi|^2 \propto \vec R^2$ is in fact proportional to the angular momentum of the skyrmion within the $(x,y)$-plane \cite{Papanicolaou91,Schuette14} so that the total density measures the total angular momentum of the string $J_z \propto \int dx_1  |\psi|^2$. In addition, the energy-momentum tensor is conserved, $\partial_\mu T_{\mu \nu} = 0$ \cite{SI}. 

We look for solitary wave solutions
\begin{align} \label{GT}
\psi(x_\mu) = \Psi(x_1 - \v x_0) e^{i (\v (x_1 - \v x_0) + \frac{ {\v}^2}{2} x_0)} e^{- i \omega x_0},
\end{align}
that can be expressed as the Galilean transform of a wave function $\Psi$ oscillating with frequency $\omega$. As the interaction $|\partial_1 \psi|^4$ breaks Galilean invariance, we expect that the function $\Psi$ will depend on the velocity $v$. The conserved densities reduce for Eq.~\eqref{GT} to functions of the variable $\c \equiv x_1 - \v x_0$ only, so that $\mathcal{T}_\nu \equiv - \v T_{0 \nu} + T_{1 \nu}$  as well as $\mathcal{J} \equiv - \v j_0 + j_1$ become independent of both space and time. In fact, the three constants are linearly dependent as one finds $\mathcal{T}_0 = (\omega - \frac{\v^2}{2}) \mathcal{J} - \v \mathcal{T}_1$. 

Moreover, for a localized solitary wave with boundary conditions $|\Psi(\c)| \to 0$ for $\c \to \pm \infty$ these constants vanish, $\mathcal{T}_\nu = 0$. Decomposing $\Psi(\c) = A(\c) e^{i \phi(\c)}$ into magnitude and phase 
we can solve these two equations for $A$ and its derivative $A'$. Remarkably, we find that they are simply parameterized by the  derivative of the phase, $\phi' = \partial_\c \phi$,
\begin{subequations}\label{eq:A-dA}
\begin{align} 
\label{DGL1}
A(\c) &= \frac{1}{|v| \sqrt{1+\s}} f_1\left(\frac{\phi'(\c)}{\v} \right),\\
\label{DGL2}
A'(\c) &= \pm 
\frac{1}{\sqrt{1+\s}}f_2\left(\frac{\phi'(\c)}{\v}, \s \right),
\end{align}
\end{subequations}
where $\s \equiv - 2\omega/v^2$ represents the frequency. The two functions are $f_1(p) = \sqrt{p(2-p)}/(1+p)$ and $f_2(p,\s) = \sqrt{p(2 \s - p + p^2)}/\sqrt{1+p}$. A closed curve in the $(A,A')$-plane, that converges to the origin for $|\c| \to \infty$, is obtained for $0 < \s < \frac{1}{8}$. This constraint yields a two-parameter family of solitary wave solutions parametrized by $\s$ and the velocity $\v$, see Ref.~\onlinecite{SI} for details.

\begin{figure*}
\includegraphics[width=\textwidth]{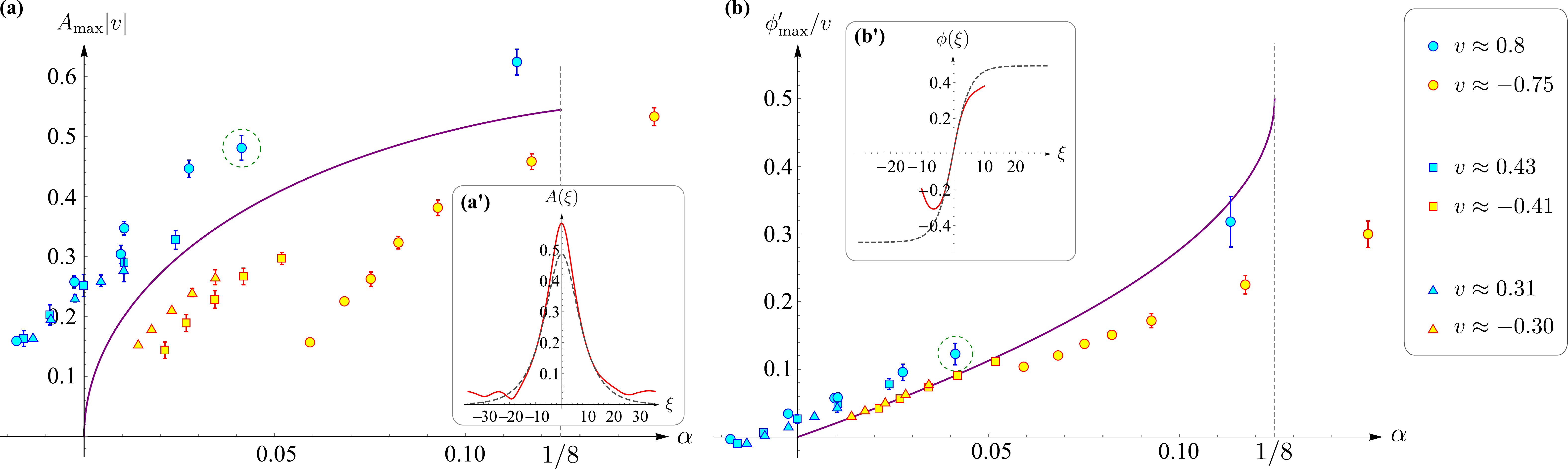}
\caption{
Profile parameters of solitary waves
extracted from micromagnetic simulations at $h = 2.16$ (symbols) for various values of dimensionless velocity $v$ (in units of $1310$ m/s). The profiles generally depend on the sign of $v$, but this non-reciprocity decreases with $|v|$ and the profiles approach the analytical predictions of Eq.~\eqref{eq:phi-A-max} (lines) valid for $v \to 0$. 
The insets show a full numerical profile (red lines) for $\alpha \approx 0.04$ (marked by dashed circles in the main panels) with a comparison to the solution of Eq.~\eqref{eq:A-dA} (dashed lines). 
}
\label{fig:4}
\end{figure*}

Taking the derivative of Eq.~\eqref{DGL1} and comparing with Eq.~\eqref{DGL2} yields a first-order ordinary differential equation for $\phi'$, that is easily solved. One obtains a wave function where both derivative $\phi'$ and magnitude  
$A$ are locally confined, see Fig.~\ref{fig:4}. Their extremal values are 
\begin{subequations}\label{eq:phi-A-max}
\begin{align}
\label{phimax}
\phi'_{\rm max} &= \frac{v}{2} (1 - \sqrt{1-8 \s}),\\
\label{Amax}
A_{\rm max} & = \frac{\sqrt{2}}{ |v|}  \frac{\sqrt{4 \s +1 - \sqrt{1-8 \s}}}{\sqrt{1+\s}(3-\sqrt{1-8 \s})},
\end{align}
\end{subequations}
and both vanish for $\s  \to 0^+$. For small $\s$ we can approximate $f_1(p) \approx \sqrt{2 p}$ and $f_2(p,\s) \approx \sqrt{p(2\s-p)}$, and we  obtain explicitly the profile for a solitary wave centered at $\c=0$ 
\begin{align} 
\label{Smalls}
\phi(\c) \approx  2 \sqrt{\s} \tanh(\sqrt{\s} \v \c),
\quad A(\c) \approx \frac{2}{ |\v| }  \frac{\sqrt{\s}}{\cosh(\sqrt{\s} \v \c))},
\end{align}
up to a constant phase shift. From these expressions we can read off the spatial width $\c_{w} \sim \frac{1}{|\v| \sqrt{\s}}$ that diverges in this limit.

Note that the amplitude  $A_{\rm max}$ in Eq.~\eqref{Amax} diverges for small velocities $\v$ but this does not invalidate the theory. The Taylor expansion of the effective potential $\mathcal{V}_{\rm eff}$ is controlled as long as the derivative, $|\partial_z\vec R| \propto |\partial_1 \psi|$, remains small. Its maximal value is attained at the center of the solitary wave, that is small $|\partial_1 \psi|^2|_{\rm max} \sim \s$, for $\s \to 0^+$.  
Moreover, non-reciprocal corrections associated with the second derivative $|\partial^2_1 \psi|/|\partial_1\psi| \sim v$ are negligible as long as the velocity is small, and, therefore, the theory \eqref{Lagrange} is under control for small $\v$ and $\s$. 
We note that non-reciprocal corrections can, in principle, be taken into account spoiling however  the simplicity of  Eqs.~\eqref{eq:A-dA}.

The solitary wave carries energy, linear and angular momentum along the string. In the limit of small $\s$, the total energy is on the order $E \sim  \frac{\sqrt{\s}}{|\v|}$, and it diverges for small velocities due to the large width of the wave, $\c_w \sim \frac{1}{|\v|\sqrt{\s}}$. The energy density, however, remains small $E/\c_w \sim \s$. For the total linear and angular momentum we obtain, respectively, $P \sim \frac{\sqrt{\s}}{|\v| \v}$ and $J_z \sim  \frac{\sqrt{\s}}{v^3}.$ Both are even more singular than $E$  due to the large amplitude, Eq.~\eqref{Amax}. As discussed above, the corresponding total currents are obtained by multiplying with $\v$.

Our micromagnetic simulations confirm the presence of stable solitary waves, see Fig.~\ref{fig:1}.  We have extracted numerically their profiles for various values of $v$ and $\alpha$, see Fig.~\ref{fig:4} and Ref.~\onlinecite{SI}. The profiles are found to be non-reciprocal, i.e., they depend on the sign of the velocity $v$. This non-reciprocity however decreases with decreasing $|v|$, and we find that they approach for $v \to 0$ the analytical predictions for this limit. As the width as well as the amplitude of the solitary wave diverges for $v \to 0$, finite-size effects of our simulation hamper the study of velocities smaller than the values listed in Fig.~\ref{fig:4}. Solitary waves are confirmed to exist for $\alpha$ within the full range between zero and $1/8$. For larger $\alpha$, the skyrmion string becomes unstable and breaks, at least, for the discretization used in our micromagnetic simulation.

Finally, we compare with the dynamics of vortex filaments in superfluids. 
Vortices are singular defects of the superfluid order parameter in contrast to skyrmions that are smooth textures. As a consequence, the dynamics of vortex filaments are  in general governed by a non-local Biot-Savart law. This non-locality is at the origin of the non-analytic dispersion, $\omega(k_z) \sim k_z^2 \log k_z$, of Kelvin waves and it leads to stretching instabilities like the development of hairpins in the filament \cite{Pismen99}. 

Only when this non-locality is neglected within the localized induction approximation (LIA), the dynamics of the vortex filament is also described by a local Schr\"odinger equation  with a Hamiltonian proportional to its length, $H = \int dx_1 \mathcal{H}$ with $\mathcal{H} = \sqrt{1+|\partial_1 \psi|^2}$ \cite{Svistunov95}. As a consequence, its length is conserved and stretching is neglected within the LIA. Interestingly, the resulting Schr\"odinger equation is integrable \cite{Konno92} and its solitons were first discussed by Hasimoto \cite{Hasimoto72}. In the limit of small intrinsic curvature $\kappa$ of the filament, $|\partial_1 \psi|$ remains small and after Taylor expanding $\mathcal{H}$ the theory approximately coincides with that of Eq.~\eqref{Lagrange} \cite{Boffetta09}. Hasimoto's soliton solution indeed reduces to Eq.~\eqref{Smalls} in this limit.

This has interesting consequences for the solitary waves studied here. Whereas the theory \eqref{Lagrange} itself is claimed to be non-integrable \cite{Boffetta09}, it is very close to an integrable theory  in the limit of its applicability. This implies that the solitary waves of the skyrmion string are expected to be approximate solitons that almost retain their shape in collisions. Using the solitary waves as carriers, the skyrmion string therefore serves as a transmission line that allows for the simultaneous, though non-reciprocal, information transfer in both directions. 
Experimentally, solitary waves could be created, e.g., by inhomogeneous spin-transfer torques \cite{Nagaosa13}.

M.G. was supported by DFG SFB1143 (Correlated Magnetism: From Frustration To Topology) Project A07 and DFG SPP2137 (Skyrmionics) Grant No.~1072/6-1. J.v.d.B is supported by SFB1143, Project A05. We acknowledge support from the UKRATOP project funded by the German Federal Ministry of Education and Research, Grant No. 01DK18002. We thank D. Sheka and Y. Gaididei for fruitful discussions, and U. Nitzsche for technical support

\bibliography{string}

\end{document}


\title{
Supplementary Materials:\\ Solitary wave excitations of skyrmion strings in chiral magnets}

\author{Volodymyr P. Kravchuk}
\affiliation{Leibniz-Institut f{\"u}r Festk{\"o}rper- und Werkstoffforschung, IFW Dresden, D-01171 Dresden, Germany}
\affiliation{Bogolyubov Institute for Theoretical Physics of National Academy of Sciences of Ukraine, 03680 Kyiv, Ukraine}

\author{Ulrich~K.~R{\"o}{\ss}ler}
\affiliation{Leibniz-Institut f{\"u}r Festk{\"o}rper- und Werkstoffforschung, IFW Dresden, D-01171 Dresden, Germany}

\author{Jeroen~van~den~Brink}
\affiliation{Leibniz-Institut f{\"u}r Festk{\"o}rper- und Werkstoffforschung, IFW Dresden, D-01171 Dresden, Germany}
\affiliation{Department of Physics, Washington University, St. Louis, MO 63130, USA}

\author{Markus Garst}
\affiliation{Institut f\"ur Theoretische Physik, Technische Universit\"at Dresden, 01062 Dresden, Germany}

\date{\today}

\begin{abstract}
Details concerning the results presented in the main text are provided. In section \ref{app:spectrum} the linear spin-wave theory is discussed that was used to compute the spectrum in Fig.~2 of the main text. In section \ref{app:cl-sw} the effective Lagrangian of Eq.~(4) in the main text is analyzed, and the derivation of the solitary wave function is further explained. Finally, section \ref{app:simuls} contains details regarding the micromagnetic simulations and explanations of the movies that are also included in the supplementary materials.
\end{abstract}

\pacs{}

\maketitle

\section{Linear spin-wave spectrum for the skyrmion string}
\label{app:spectrum}

The evaluation of the linear spin-wave spectrum for the skyrmion string is based on a standard technique that was 
previously applied to a variety of two-dimensional magnetic textures~\cite{Ivanov96,Ivanov99,Sheka01,Schuette14}. Very recently, this spectrum was also computed for the skyrmion string by  Lin {\it et al.} \cite{Lin2019} with results in complete agreement with ours. Here, we review the method, and we give details of the eigenvalue problem that has to be solved. 

The Landau-Lifshitz equation is the Euler-Lagrange equation of the Lagrangian density
\begin{equation}\label{eq:lagrangian}
L=\frac{M_s}{\gamma}\vec{A}(\vec{n})\cdot\partial_t\vec{n}-\mathcal{V},
\end{equation}
where $\vec{n}$ is a unit vector field describing the orientation of the magnetization,
$\mathcal{V}$ is the potential density, and 
$\vec{A}(\vec{n})$ is the spin-gauge field that obeys $\frac{\partial A_j}{\partial n_i}-\frac{\partial A_i}{\partial n_j}=\epsilon_{ijk}n_k$. Let $\vec{n}_0(\vec{r})$ be a static equilibrium solution of the magnetization. Small deviations from the equilibrium state can be parameterized by means of a complex valued function $\varphi(\vec{r},t)$ in the following way
\begin{equation}\label{eq:psi-param}
\vec{n}=\vec{n}_0\sqrt{1-2|\varphi|^2}+\vec{e}^+\varphi+\vec{e}^-\varphi^*,
\end{equation}
where $\vec{e}^{\pm}=(\vec{e}_1\pm \mathrm{i}\vec{e}_2)/\sqrt{2}$ with $\vec{e}_1\times\vec{e}_2=\vec{n}_0$. This is a particular case of a general representation of spin-operators  discussed in Ref.~\onlinecite{Tyablikov75}. Using $\vec{A}(\vec{n})=\vec{A}(\vec{n}_0)+\varphi\frac{\partial\vec{A}}{\partial n_i} e^+_i+\varphi^*\frac{\partial\vec{A}}{\partial n_i} e^-_i+O(|\varphi|^2)$ and the properties  $\vec{e}^+\cdot\vec{e}^+=\vec{e}^-\cdot\vec{e}^-=0$, $\vec{e}^+\cdot\vec{e}^-=1$ the harmonic part of the Lagrangian density \eqref{eq:lagrangian} is obtained in the form
\begin{equation}\label{eq:L-harm}
L^{(2)}=\frac{\mathrm{i}}{2}\frac{M_s}{\gamma}\vec{\varPhi}^\dag\tau^z\partial_t\vec{\varPhi}-\mathcal{V}^{(2)},
\end{equation}
where $\vec{\varPhi}=(\varphi,\varphi^*)^\textsc{t}$, $\tau^z$ is the third Pauli matrix and $\mathcal{V}^{(2)}$ is the harmonic part of the energy density. 

In the following we consider the potential for cubic chiral magnets given in Eq.~(1) of the main text. Using dimensionless time $t'=t\omega_{c2}$ and space $\vec{r}'=\vec{r}Q$ coordinates results in the following harmonic Lagrangian \begin{equation}\label{eq:L-harm-dmls}
\mathscr{L}^{(2)}=\frac12 \vec{\varPhi}^\dag\left(\mathrm{i}\tau^z\partial_{t'}-\mathcal{H}\right)\vec{\varPhi},\quad\mathcal{H}=\begin{pmatrix}
\mathscr{H}^+ & \mathscr{W}^- \\
\mathscr{W}^+ & \mathscr{H}^-
\end{pmatrix}.
\end{equation}
Here $\mathscr{H}^\pm=-\Delta-\mathscr{V}_0-2\mathscr{V}^\pm$ where $\Delta$ is the Laplace operator and
\begin{equation}\label{eq:potentials}
\begin{split}
\mathscr{V}_0=&-\vec{n}_0\cdot\Delta\vec{n}_0+2\,\vec{n}_0\!\cdot\![\vec{\nabla}\times\vec{n}_0]-h(\vec{n}_0\cdot\hat{\vec z})\\
&+\mathrm{Re}(\vec{e}^+\cdot\Delta\vec{e}^-)-2\mathrm{Re}[\vec{e}^+\cdot(\vec{\nabla}\times\vec{e}^-)],\\
\mathscr{V}^\pm=&e^\mp_i(\vec{\nabla}e^\pm_i\cdot\vec{\nabla})-\left([\vec{e}^\pm\times\vec{e}^\mp]\cdot\vec{\nabla}\right),\\
\mathscr{W}^\pm=&-(\vec{e}^\pm\cdot\Delta\vec{e}^\pm)+2\vec{e}^\pm\cdot[\vec{\nabla}\times\vec{e}^\pm].
\end{split}
\end{equation}
It is convenient to utilize the constraint $|\vec{n}_0|=1$ by means of the angular parameterization $\vec{n}_0=\sin\theta\vec{\varepsilon}+\cos\theta\hat{\vec z}$, with $\vec{\varepsilon}=\cos\phi\hat{\vec x}+\sin\phi\hat{\vec y}$. In this case vectors $\vec{e}_1$ and $\vec{e}_2$ can be defined as follows $\vec{e}_1=\partial_{\phi}\vec{\varepsilon}$, $\vec{e}_2=-\partial_{\theta}\vec{n}_0$ and the potentials \eqref{eq:potentials} obtain the form
\begin{equation}\label{eq:V-W}
\begin{split}
\mathscr{V}_0=&\frac{(\vec{\nabla}\theta)^2}{2}+(1-3\cos^2\theta)\left[\frac{(\vec{\nabla}\phi)^2}{2}-\partial_z\phi\right]\\
&+(\vec{\nabla}\theta\times\vec{\varepsilon})_z+3\sin\theta\cos\theta(\vec{\varepsilon}\cdot\vec{\nabla}\phi)-h\cos\theta,\\
\mathscr{V}^\pm=&\pm\mathrm{i}\left[\left(\vec{n}_0-\cos\theta\vec{\nabla}\phi\right)\cdot\vec{\nabla}\right],\\
\mathscr{W}^{\pm}=&-\frac{(\vec{\nabla}\theta)^2}{2}+\sin^2\theta\left[\frac{(\vec{\nabla}\phi)^2}{2}-\partial_z\phi\right]+(\vec{\varepsilon}\times\vec{\nabla}\theta)_z\\
&+\sin\theta\cos\theta(\vec{\varepsilon}\cdot\vec{\nabla}\phi)\mp\mathrm{i}\vec{\nabla}\cdot\vec{n}_0\\
&\pm\mathrm{i}\sin\theta\left[2(\vec{\varepsilon}\times\vec{\nabla}\phi)_z-(\vec{\nabla}\theta\cdot\vec{\nabla}\phi)\right].
\end{split}
\end{equation}

Note that Eqs.~\eqref{eq:L-harm-dmls} and \eqref{eq:V-W} determine  the linear dynamics for an arbitrary equilibrium state $\vec{n}_0$. Now, let us  consider $\vec{n}_0$ in the form of a skyrmion solution. In the cylindrical frame of reference $(\varrho,\chi,z)$ the model (1) of the main text permits the separation of variables for the polar angle $\theta=\theta(\varrho)$ and the azimuthal angle $\phi=\phi(\chi)$. The function $\theta(\varrho)$ is determined by the equation \cite{Bogdanov94}
\begin{equation}\label{eq:theta}
\begin{split}
&\Delta_\varrho\theta-\frac{\sin\theta\cos\theta}{\varrho}+2\frac{\sin^2\theta}{\varrho^2}-h\sin\theta=0,
\end{split}
\end{equation}
 where $\phi=\chi+\mathrm{sgn}(D)\pi/2$, and $\Delta_\varrho$ denotes the radial part of the Laplacian operator. The boundary conditions $\theta(0)=\pi$ and $\theta(\infty)=0$ applied to Eq.~\eqref{eq:theta} results in a skyrmion solution. In this case the potentials \eqref{eq:V-W} have the form
 \begin{equation}\label{eq:V-W-skr}
 \begin{split}
 \mathscr{V}_0=&\frac{(\partial_\varrho\theta)^2}{2}+\frac{1-3\cos^2\theta}{2\varrho^2}+\partial_\varrho\theta+3\frac{\sin\theta\cos\theta}{\varrho}-h\cos\theta,\\
 \mathscr{V}^\pm=&\pm\mathrm{i}\left(\frac{\sin\theta}{\varrho}-\frac{\cos\theta}{\varrho^2}\right)\partial_{\chi}\pm\mathrm{i}\cos\theta\partial_{z},\\
 \mathscr{W}^\pm=&\mathscr{W}=-\frac{(\partial_\varrho\theta)^2}{2}+\frac{\sin^2\theta}{2\varrho^2}-\partial_\varrho\theta+\frac{\sin\theta\cos\theta}{\varrho}.
 \end{split}
 \end{equation}
Applying the Fourier transform 
\begin{equation}\label{eq:FT}
\begin{split}
&\bar\varphi_{k_zm\omega}(\varrho)=\!\!\int\limits_{-\infty}^{\infty}\!\!\mathrm{d}t\!\!\int\limits_{-\infty}^{\infty}\!\!\mathrm{d}z\!\!\int\limits_{0}^{2\pi}\mathrm{d}\chi e^{\mathrm{i}(\omega t-k_zz-m\chi)}\varphi(\varrho,\chi,z,t),\\
&\varphi=\frac{1}{(2\pi)^{3/2}}\!\!\int\limits_{-\infty}^{\infty}\!\!\!\mathrm{d}\omega\!\!\!\int\limits_{-\infty}^{\infty}\!\!\!\mathrm{d}k_z\!\!\!\sum\limits_{m=-\infty}^{\infty}\!\!e^{-\mathrm{i}(\omega t-k_zz-m\chi)}\bar\varphi_{k_zm\omega}(\varrho)
\end{split}
\end{equation}
to the equations of motion generated by the Lagrangian~\eqref{eq:L-harm-dmls} one obtains the following eigenvalue problem
\begin{equation}\label{eq:EVP}
\mathbb{H}_{k_zm}\tilde{\vec{\varPhi}}_{k_zm\omega}=\omega\tau^z\tilde{\vec{\varPhi}}_{k_zm\omega},\quad\mathbb{H}_{k_zm}=\begin{pmatrix}
\mathscr{H}_{k_zm}&\mathscr{W}\\
\mathscr{W}&\mathscr{H}_{-k_z-m}
\end{pmatrix}.
\end{equation}
Here $\tilde{\vec{\varPhi}}_{k_zm\omega}=(\bar{\varphi}_{k_zm\omega},\bar{\varphi}^*_{-k_z-m-\omega})^\textsc{t}$ and $\mathscr{H}_{k_zm}=-\Delta_\varrho+\frac{m^2}{\varrho^2}+k_z^2-\mathscr{V}_0+2k_z\cos\theta+2m\left(\frac{\sin\theta}{\varrho}-\frac{\cos\theta}{\varrho^2}\right)$ with $\mathscr{V}_0$ and $\mathscr{W}$ defined in Eq.~\eqref{eq:V-W-skr}. The sign of the azimuthal quantum number $m$ defined here coincides with the definition of  Ref.~\onlinecite{Schuette14} but it is opposite to the one defined in Ref.~\onlinecite{Kravchuk18}.

\begin{figure}
	\includegraphics[width=\columnwidth]{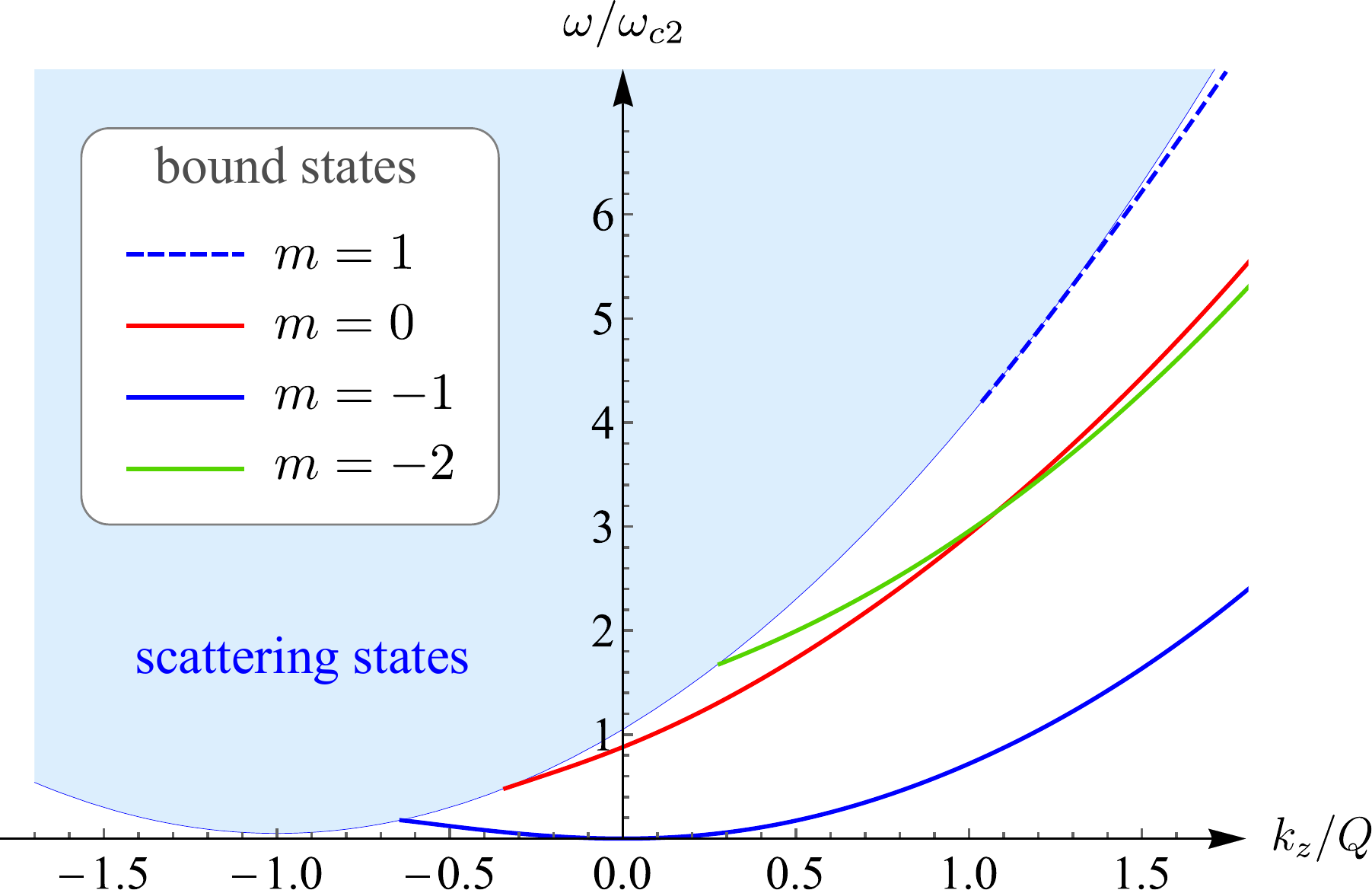}
	\caption{Linear spin-wave spectrum of the skyrmion string for a field $h=1.05$.}\label{fig:spct}
\end{figure}

For each given value of the dimensionless field $h>1$ the linear analysis proceeds in two steps: $(i)$ by means of numerically solving Eq.~\eqref{eq:theta} we obtain the skyrmion profile $\theta(\varrho)$; $(ii)$ using the profile $\theta(\varrho)$ we determine the operator $\mathbb{H}_{k_zm}$ and numerically solve the eigenvalue problem \eqref{eq:EVP}. An example of the resulting energy spectrum for $h=2.16$ is shown in Fig.~2(a) of the main text with two magnon-skyrmion bound states within the given range of wavevectors: the breathing mode with $m=0$ and the translational mode with $m=-1$. Additional bound modes appear for smaller values of $h$. In Fig.~\ref{fig:spct} the spectrum is shown for $h=1.05$ with an additional high-frequency gyrotropic mode with $m=1$ and a quadrupolar mode with $m=-2$.

\section{Solitary wave solution for the skyrmion string}
\label{app:cl-sw}

The solitary wave solutions of the effective Lagrangian in Eq.~(4) of the main text is discussed in further detail. 
We start with the corresponding Euler-Lagrange equations of motion that read
\begin{subequations}\label{eq:eqs}
	\begin{align}
\label{eq:eqs1}-\mathrm{i}\partial_{0}\psi=&\frac12\partial_1^2\psi-\frac14\partial_1\left(|\partial_1\psi|^2\partial_1\psi\right),\\
\label{eq:eqs2}\mathrm{i}\partial_{0}\psi^*=&\frac12\partial_1^2\psi^*-\frac14\partial_1\left(|\partial_1\psi|^2\partial_1\psi^*\right).
	\end{align}
\end{subequations}
Multiplication of Eq.~\eqref{eq:eqs1} by $\psi^*$ and Eq.~\eqref{eq:eqs2} by $-\psi$
and subsequent summation yields the conservation law 
\begin{equation}\label{eq:j-div}
\partial_\mu j_\mu = 0,
\end{equation}
with the density $j_0$ and the associated current $j_1$ given by 
\begin{subequations}\label{eq:j}
	\begin{align}
		\label{eq:j0}j_0=&|\psi|^2,\\
		\label{eq:j1}j_1=&\frac{\mathrm{i}}{2}(\psi\partial_1\psi^*-\psi^*\partial_1\psi)\left(1-\frac12|\partial_1\psi|^2\right).
	\end{align}
\end{subequations}
For appropriate boundary conditions the total number of particles $N=\int|\psi|^2\mathrm{d}x_1$ is constant in time. 
Moreover, the symmetry with respect to space-time translations leads to the conservation law
\begin{equation}\label{eq:T-cons}
\partial_\mu T_{\mu\nu}=0
\end{equation}
for the energy-momentum tensor 
\begin{align}
	T_{\mu \nu} = - \delta_{\mu\nu} \mathcal{L} + \frac{\partial \mathcal{L}}{\partial \partial_\mu \psi} \partial_\nu \psi 
	+ \frac{\partial \mathcal{L}}{\partial \partial_\mu \psi^*} \partial_\nu \psi^*.
\end{align}
Consequently, the total energy $E=\int T_{00}\mathrm{d}x_1$ and total linear momentum $P=-\int T_{01}\mathrm{d}x_1$ are constant in time. The components of the energy-momentum tensor read explicity
\begin{equation}\label{eq:T}
\begin{split}
T_{00}&=\frac12|\partial_1\psi|^2-\frac18|\partial_1\psi|^4,\\
T_{01}&=\frac{\mathrm{i}}{2}(\psi^*\partial_1\psi-\psi\partial_1\psi^*),\\
T_{10}&=-\frac{\mathrm{1}}{2}(\partial_0\psi^*\partial_1\psi+\partial_0\psi\partial_1\psi^*)\left(1-\frac12|\partial_1\psi|^2\right),\\
T_{11}&=-\frac{\mathrm{i}}{2}(\psi^*\partial_0\psi-\psi\partial_0\psi^*)-\frac12|\partial_1\psi|^2+\frac38|\partial_1\psi|^4.
\end{split}
\end{equation}
For the Ansatz of the solitary wave function given in Eq.~(5) of the main text with velocity $v$, all these components $T_{\mu\nu}$ as well as $j_\mu$ are reduced to functions of the variable $\xi=x_1-v x_0$. The conservation laws \eqref{eq:j-div} and \eqref{eq:T-cons} then imply that the following quantities are independent of space and time,
\begin{subequations}\label{eq:J-T}
	\begin{align}
\mathcal{J}=&-vj_0+j_1=A^2\left[\phi'-(\phi'+v)\frac{|\psi'|^2}{2}\right],\\
\mathcal{T}_0=&-vT_{00}+T_{10}=\frac{v}{2}|\psi'|^2\left(1-\frac34|\psi'|^2\right)\\ \nonumber
&+A^2\left(\omega-\frac{v^2}{2}\right)(\phi'+v)\left(1-\frac{|\psi'|^2}{2}\right),\\
\mathcal{T}_1=&-vT_{01}+T_{11}=\\ \nonumber
&-\frac{|\psi'|^2}{2}\left(1-\frac34|\psi'|^2\right)
-A^2\left(\omega-\frac{v^2}{2}\right),
	\end{align}
\end{subequations}
where $|\psi'|^2=A'^2+A^2(\phi'+v)^2$. These quantities are linearly dependent $\mathcal{T}_0+v\mathcal{T}_1=(\omega-\frac{v^2}{2})\mathcal{J}$.

\begin{figure}
\includegraphics[width=\columnwidth]{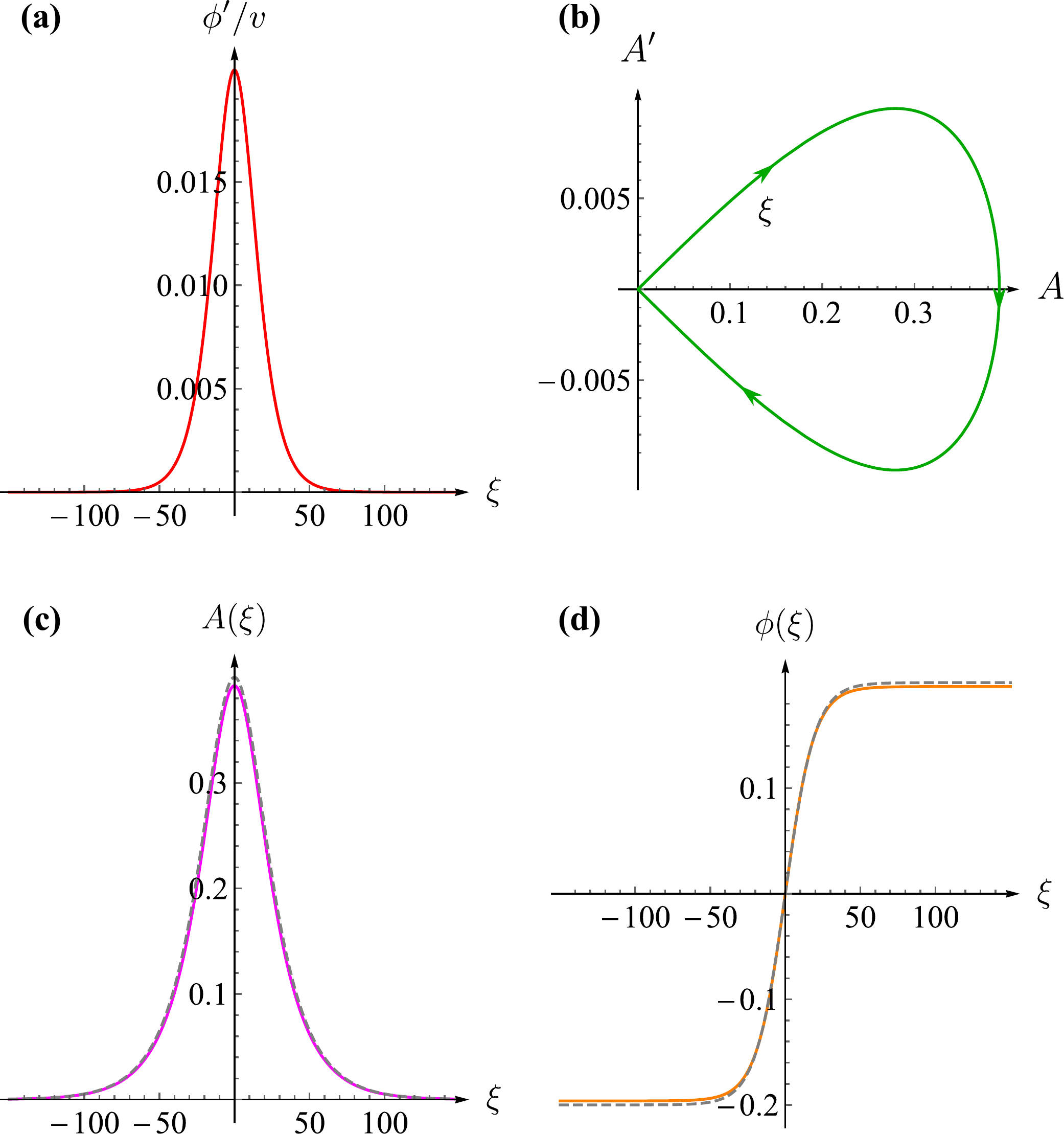}
\caption{Solitary wave of the skyrmion string with velocity $v=0.5$ and parameter $\alpha=0.01$. (a) Solution for the derivative of the phase, $\phi'(\xi) = v \kappa(\xi)$, obtained by solving Eq.~\eqref{eq:kappa}. (b) The corresponding trajectory in the $(A,A')$-plane using Eq. (6) of the main text. (c) The magnitude and (d) the phase of the solitary wave. The dashed lines in panel (c) and (d) show the low-amplitude approximations given in Eq.~(8) of the main text. 
}\label{fig:kappa}
\end{figure}

In case of a spatially localized solution with boundary conditions $A(\pm\infty)=0$ and $A'(\pm\infty)=0$ these constants vanish $\mathcal{J}=\mathcal{T}_1=\mathcal{T}_2=0$. 
Solving now any two equations of \eqref{eq:J-T} for $A$ and $A'$ one obtains Eqs.~(6) of the main text. The resulting ordinary differential equation for the derivative of the phase $\phi'(\xi) = v \kappa(\xi)$ reads in terms of the auxiliary localized function $\kappa(\xi)$ centered at $\xi = 0$
\begin{subequations}\label{eq:kappa}
	\begin{align}
\label{eq:kappa-eq}&\kappa'=- \text{sgn}(\xi)|v|\frac{\kappa(1+\kappa)^{3/2}\sqrt{(2-\kappa)(2\alpha-\kappa+\kappa^2)}}{1-2\kappa},\\
\label{eq:kappa-ini}&\kappa(0)=(1-\sqrt{1-8\alpha})/2.
\end{align}
\end{subequations} 
where $\alpha = -2 \omega/v^2$.
The initial condition \eqref{eq:kappa-ini} originates from the condition $A'(0)=0$ that the derivative of the magnitude vanishes at the center of the solitary wave. Note that the initial condition is formulated for the point $\xi = 0$, which is exactly at the boundary of the definitional domain of the right-hand side of  Eq.~\eqref{eq:kappa-eq}. The assumptions of the Peano existence theorem \cite{Peano1890} are therefore not fulfilled, and the Cauchy differential problem \eqref{eq:kappa} can have more than one solution. As a consequence, in addition to the trivial solution $\kappa(\xi)\equiv\kappa(0)=\text{const}$, one obtains the localized one with $\kappa(\pm\infty)=0$, see Fig.~\ref{fig:kappa}(a). Substitution of $\phi'=v\kappa$ into Eq.~(6) of the main text results in the phase trajectory shown in Fig.~\ref{fig:kappa}(b) and the  magnitude shown in Fig.~\ref{fig:kappa}(c). The phase kink shown in Fig.~\ref{fig:kappa}(d) is obtained by computing the integral $\phi(\xi)=v\int_0^\xi\kappa(\xi')\mathrm{d}\xi'$.

In the limit $\alpha \to 0^+$, the total energy of the solitary wave is given by $E\approx\frac{4\sqrt{\alpha}}{|v|}$, the total momentum reads $P\approx \frac{8\sqrt{\alpha}}{v|v|}$ and the total number of particles is $N\approx 8\sqrt{\alpha}/v^3$. The latter quantifies the total angular momentum of the skyrmion string, $J_z \propto N$.

\section{Micromagnetic simulations}\label{app:simuls}

\begin{figure*}
\includegraphics[width=\textwidth]{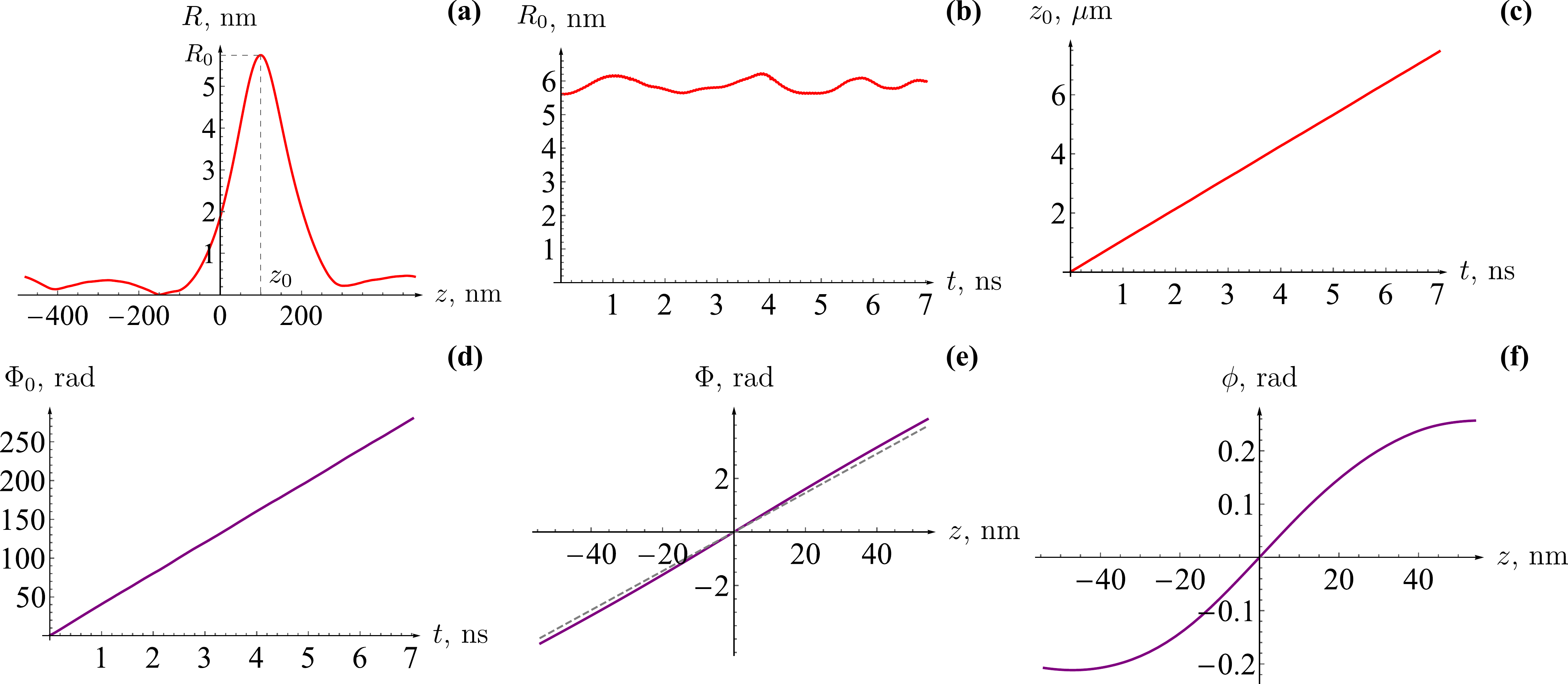}
\caption{Numerically extracted profile of the solitary wave for the example given in Fig.~1 of the main text with positive velocity $v > 0$. (a) Spatial distribution of the magnitude $R=|X+\mathrm{i}Y|$ at time $t=6.4$~ns. It shows a localized excitation with amplitude $R_0$ and position $z_0$ that both evolve in time as shown in panels (b) and (c), respectively.
(d) Time evolution of the phase $\varPhi_0=\text{arg}(X_0+\mathrm{i}Y_0)$ at the center of the solitary wave, $(X_0,Y_0)=(X(z_0),Y(z_0))$. (e) Spatial distribution of the phase within the comoving frame of reference $\varPhi(z)=\text{arg}[X(z-z_0)+\mathrm{i}Y(z-z_0)]-\varPhi_0$ within the region $|z-z_0|<\xi_w$, where $\xi_w$ is the width of the solitary wave, at time $t=6.4$~ns. The dashed line is given by $k_z^{\rm sw} z$ with $k_z^{\rm sw} =QV/V_0$, see text, and the difference $\phi(z)=\varPhi(z)-k_z^{\rm sw} z$ is shown in panel (f).}
	\label{fig:soliton-detailes}
\end{figure*}

The micromagnetic simulations were performed with the OOMMF code \cite{OOMMF} supplemented with the extension for the Dzyaloshinskii-Moriya interaction (DMI) in cubic crystals  \cite{Cortes-Ortuno18}. We used the parameters for the cubic chiral magnet FeGe with four Fe and four Ge atoms per unit cell that were given in Ref.~\onlinecite{Beg15}: exchange constant $A=8.78\times10^{-12}$ J/m, saturation magnetization $M_s=3.84\times10^{5}$ A/m, and DMI constant $D=1.58\times10^{-3}$ J/m$^2$. Magnetic field $\mu_0H=0.8$~T was fixed for all simulations. The magnetostatic interaction was neglected. These material parameters determine the length scale $Q^{-1}\approx 11.1$ nm and the time scale $\omega_{c2}^{-1}\approx15.3$ ps.

We simulated the magnetization dynamics within a cylinderically shaped sample oriented along the field with a diameter 100 nm and the length was varied from 0.8 to 5~$\mu$m depending on the width of the solitary wave. For most of the cases the spatial discretization was fixed to $(\Delta x,\Delta y, \Delta z)=(1\,\text{nm},1\,\text{nm},3\,\text{nm})$. This enabled us to consider  skyrmion string excitations with deformation gradients up to $|\partial_z\vec{R}|_\text{max}\approx 0.25$. For larger deformations we observed that the string broke producing a pair of oppositely charged Bloch points that quickly separated. For the largest samples the spatial discretization was increased to $(\Delta x,\Delta y, \Delta z)=(2\,\text{nm},2\,\text{nm},5\,\text{nm})$.

Periodic boundary conditions were applied to the top and bottom surfaces of the cylinder so that the skyrmion string is effectively forming a loop. At the side surfaces of the cylinder we employed Dirichlet boundary conditions with a magnetization pointing along the field,
$\vec{n}=\hat{\vec z}$.  Open boundary conditions on the side surfaces would have  the disadvantage that the magnetization twists at the boundary \cite{Rohart13} effectively reducing the lateral size of the sample.

Each simulation was performed in two steps: $(i)$ an initial magnetization $\vec{n}_{\text{ini}}(\vec{r})$ was generated containing a solitary wave of a form close to the one expected theoretically. This magnetization was relaxed using the Landau-Lifshitz-Gilbert equation with a large damping parameter $\alpha_\textsc{g}=0.5$ for several tens of picoseconds. $(ii)$ Afterwards the damping was switched off, $\alpha_\textsc{g}=0$, and the simulation was continued for a total time which varied from 7 to 40 ns depending on the  velocity of the solitary wave. 

\begin{figure*}
	\includegraphics[width=\textwidth]{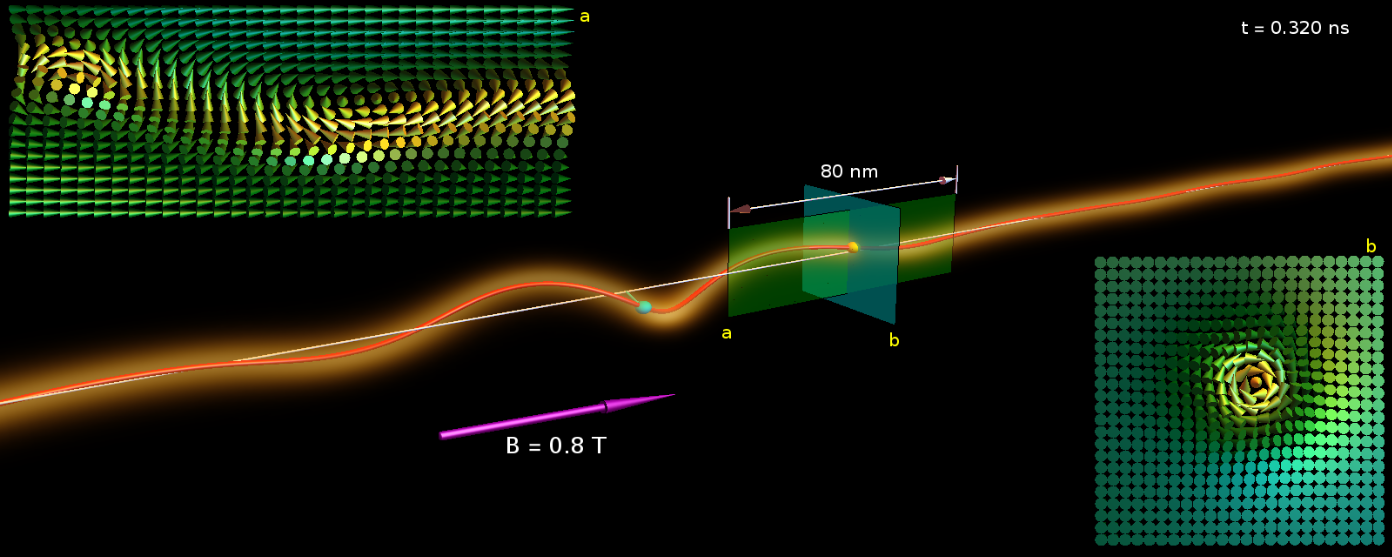}
	\caption{Snapshot of the movie {\fontfamily{qcr}\selectfont movie2.mkv} showing the dynamics of the solitary wave, whose parameters are marked by the superscript $^{\rm m}$ in Table~\ref{tbl:raw-data}. The red line shows the position of the skyrmion string as defined by Eq.~(2) of the main text. The color density represents the distribution of the topological charge $\rho_{\rm top}$. Panels `a' and `b' show the distribution of the magnetization along the cross sections determined by the planes `a' and `b', respectively. The cyan point has coordinates $(X_0,Y_0,z_0)$ and shows the position of the solitary-wave maximum. The yellow point has coordinates $(X(0),Y(0),0)$ and shows the intersection of the string and the plane `b'. Note the opposite sense of rotation of the cyan and yellow points in the movie. This follows directly from the structure of the Galilean transformation of the wavefunction in Eq.~(5) of the main text. Whereas the cyan point in the comoving frame of reference is expected to rotate with the dimensionless frequency $\omega -v^2/2 =  - (1 + \alpha)v^2/2 <0$ that is negative, the yellow point in the fixed frame of reference instead rotates on average with a positive frequency $\omega+ v^2/2  =   (1 - \alpha)v^2/2 > 0$ for $0 < \alpha < 1/8$. No  scalings of length are applied here.
	}\label{fig:movie}
\end{figure*}

\subsection{Determination of the solitary wave profile}

During the second stage $(ii)$ of the simulation the magnetization snapshots were saved at timesteps $\Delta t=10$~ps. From each snapshot we extracted the skyrmion string coordinates $X$ and $Y$ by means of Eq.~(2) of the main text, with which we numerically constructed the function $\Psi_{\rm sim}(z,t)=X(z,t)+\mathrm{i}Y(z,t)$. A typical profile of the solitary wave defined by the magnitude $R(z,t)=|\Psi_{\rm sim}(z,t)|$ as a function of $z$ at a certain fixed time $t$  is shown in Fig.~\ref{fig:soliton-detailes}(a). From the profile function $R(z,t)$ we extracted three time-dependent quantities: the amplitude $R_0(t)=\max_zR(z,t)$, the position $z_0(t):\,R_0(t)=R(z_0(t),t)$, and the width $\xi_w(t)$ by means of fitting $R(z,t)$ with the trial function $R^{\rm fit}(z)=R_0/\cosh(\frac{z-z_0}{\xi_w})$ at each moment of time. Results for $R_0(t)$ and $z_0(t)$ are shown in Fig.~\ref{fig:soliton-detailes}(b) and (c), respectively. Due to parasitic magnon excitations the amplitude $R_0(t)$ weakly fluctuates around a time-independent constant value $\langle R_0\rangle_t$. The time averaged amplitude $\langle R_0\rangle_t$ and width  $\langle \xi_w \rangle_t$ are listed in Table~\ref{tbl:raw-data}. 

Moreover, fitting the dependence $z_0(t)$ by a linear function $z_0^{\rm fit}(t)=Vt+z_0^{\rm ini}$, we obtain the velocity $V$, that is also listed in Table~\ref{tbl:raw-data}. In order to determine the rotation frequency $\Omega$ of the solitary wave within the comoving frame of reference, we considered the phase $\varPhi_0(t)=\text{arg}[X_0(t)+\mathrm{i}Y_0(t)]$ of the solitary wave at its center, where $X_0(t)=X(z_0(t),t)$ and $Y_0(t)=Y(z_0(t),t)$, see Fig.~\ref{fig:soliton-detailes}(d). Fitting the time dependence $\varPhi_0(t)$ by the linear function $\varPhi_0^{\rm fit}(t)=\Omega t+\varPhi_0^{\rm ini}$ yields the frequency $\Omega$, see Table~\ref{tbl:raw-data}. In addition, we determined the phase within the comoving frame of reference $\varPhi(z)=\text{arg}[X(z-z_0)+\mathrm{i}Y(z-z_0)]-\varPhi_0$, see Fig.~\ref{fig:soliton-detailes}(e). Due to parasitic magnon excitations, the phase $\varPhi(z)$ exhibits fluctuations, which are small within the region of the solitary wave $|z-z_0|<\xi_w$, but they are significant outside this region $|z-z_0|>\xi_w$ (not shown). The derivative of the phase at the center of the solitary wave averaged over time $\langle\varPhi'(0)\rangle_t$ is listed in Table~\ref{tbl:raw-data}.

With this procedure we constructed the solitary wave function
\begin{equation}\label{eq:Psi_sim}
\Psi_{\rm sim}(z,t)=R(z-Vt)e^{\mathrm{i}\varPhi(z-Vt)+\mathrm{i}\Omega t}.
\end{equation}
Comparison of Eq.~\eqref{eq:Psi_sim} with Eq.~(5) of the main text allows to connect the numerical data with our analytical results. The magnitude function $A$ is related to $R$ via $A =R Q \sqrt{2 b_1/a_1}$ with the coefficients $a_1$ and $b_1$ defined in Fig.~3 of the main text. The dimensionless velocity is given by $v = V/V_0$ where $V_0 = \Omega_0/Q \approx 1308$ m/s, and $\Omega_0=2 a_1 \omega_{c2}\approx 1.177\times10^{11}$~rad/s. With the help of the rotation frequency $\Omega$ and the velocity $V$ we can extract the dimensionless frequency $\omega  = \frac{V^2}{2V_0^2} - \Omega/\Omega_0$, from which follows the dimensionless parameter $\alpha = - 2 \omega/v^2=2\frac{\Omega V_0^2}{\Omega_0V^2}-1$ that is a central quantity in the theoretical description, see Fig. (4) of the main text.  

The solitary wave phase profile $\phi(z)$ is determined as $\phi(z)=\varPhi(z)-k_z^{\rm sw} z$, where $k_z^{\rm sw} =QV/V_0$ can be interpreted as a wavevector associated with the solitary wave.

\subsection{Movies of the solitary wave}

For better illustration of the solitary wave dynamics we provide three movies constructed from our micromagnetic simulations.

{\fontfamily{qcr}\selectfont movie1.mkv} shows the full time evolution of the snapshots given in Fig.~1 of the main text. Shown are 3 ns of the dynamics of two solitary waves moving in opposite directions. The length of the cylinder is 0.96~$\mu$m with periodic boundary conditions at the top and the bottom. Lengths in the $(x,y)$-plane are upscaled by a factor of five as in Fig.~1 of the main text. The vertical arrow indicates the direction of the magnetic field $\vec{B}=\hat{\vec z}$(0.8~T). 

{\fontfamily{qcr}\selectfont movie2.mkv} illustrates the solitary wave with the largest amplitude of all simulated waves. 
The corresponding parameters are marked by the superscript $^{\rm m}$ in Table~\ref{tbl:raw-data}. Shown are 
0.8~ns of its dynamics; details are explained in Fig.~\ref{fig:movie}.

{\fontfamily{qcr}\selectfont movie3.mkv} is an alternative representation of the dynamics shown in {\fontfamily{qcr}\selectfont movie2.mkv}. The local gyrovector field $\vec{\rho}=\hat{\vec x}_i\epsilon_{ijk}\vec{n}\cdot\left[\partial_{x_j}\vec{n}\times\partial_{x_k}\vec{n}\right]/(8\pi)$ was evaluated from the magnetization field, and it is shown by the green arrows in addition to the red line that represents Eq.~(2) of the main text as in the previous two movies. The spatial distribution of the quantity $|\vec{\rho}|$ is represented both by the length of the green arrows as well as by the color density. Note that the vector $\vec{\rho}$ is tangential to the string as defined by the red line.

\begin{table*}[h]
		\renewcommand{\arraystretch}{1.5}	
	\begin{tabular}{>{\RaggedRight\arraybackslash}p{3.75cm}>{\RaggedRight\arraybackslash}p{3.75cm}>{\RaggedRight\arraybackslash}p{2.25cm}>{\RaggedRight\arraybackslash}p{2.25cm}>{\RaggedRight\arraybackslash}p{3cm}}\hline\hline
		Velocity $V$,~km/s & Frequency of rotation in the comoving r.f. $\Omega$,~GHz & Amplitude $R_0$,~nm & Half-width $\xi_w$,~nm & Derivative of the phase at the solitary-wave center  $\varPhi'(0)$,~nm$^{-1}$ \\  \hline
		$^{\rm m}0.93286\pm6.3\times10^{-5}$ & $5.305\pm7.4\times10^{-4}$ & $9.6\pm0.3$ & $37\pm2$ &  $0.0846\pm0.0024$\\
		$1.00468\pm8.1\times10^{-5}$ & $5.754\pm8.2\times10^{-4}$ & $6.8\pm0.3$ & $53\pm4$ &  $0.0776\pm0.0011$\\
		$1.06104\pm1.0\times10^{-4}$ & $6.331\pm1.2\times10^{-3}$ & $6.0\pm0.2$ & $54\pm2$ &  $0.07998\pm0.0009$\\
		$1.08760\pm9.5\times10^{-5}$ & $6.544\pm1.1\times10^{-3}$ & $4.6\pm0.15$ & $66\pm5$ &  $0.07910\pm0.0006$\\
		$1.09221\pm7.4\times10^{-5}$ & $6.594\pm8.0\times10^{-4}$ & $3.97\pm0.2$ & $71\pm11$ &  $0.07948\pm0.0005$\\		
		$1.10592\pm9.5\times10^{-5}$ & $6.679\pm1.0\times10^{-3}$ & $3.33\pm0.13$ & $82\pm7$ &  $0.07870\pm0.00032$\\	
		$1.13773\pm3.7\times10^{-4}$ & $6.962\pm4.4\times10^{-3}$ & $2.0\pm0.05$ & $152\pm3$ &  $0.07800\pm0.00032$\\	
		$-0.87884\pm7.3\times10^{-5}$ & $4.8607\pm9.2\times10^{-4}$ & $8.7\pm0.2$ & $45\pm2$ &  $-0.07858\pm0.0012$\\
		$-0.93909\pm7.0\times10^{-5}$ & $5.3946\pm8.5\times10^{-4}$ & $7.0\pm0.2$ & $51\pm2$ &  $-0.07915\pm0.0009$\\	
		$-0.97287\pm1.0\times10^{-4}$ & $5.6623\pm1.2\times10^{-3}$ & $5.6\pm0.2$ & $62\pm4$ &  $-0.07844\pm0.0007$\\			
		$-0.99312\pm6.6\times10^{-5}$ & $5.8445\pm6.2\times10^{-4}$ & $4.6\pm0.15$ & $69\pm6$ & $-0.07864\pm0.0004$\\		
		$-1.00521\pm8.7\times10^{-5}$ & $5.9478\pm9.8\times10^{-4}$ & $3.7\pm0.17$ & $82\pm9$ &  $-0.07868\pm0.0004$\\		
		$-1.01561\pm9.3\times10^{-5}$ & $6.0325\pm9.2\times10^{-4}$ & $3.17\pm0.07$ & $90\pm4$ &  $-0.07831\pm0.0002$\\			
		$-1.02910\pm2.2\times10^{-4}$ & $6.1414\pm2.7\times10^{-4}$ & $2.18\pm0.05$ & $137\pm7$ &  $-0.07814\pm0.0002$\\
		\hline
		$0.54606\pm1.6\times10^{-5}$ & $1.6716\pm7.0\times10^{-5}$ & $8.6\pm0.4$ & $122\pm13$ &  $0.04051\pm0.0003$\\
		$0.55616\pm1.6\times10^{-5}$ & $1.7112\pm8.0\times10^{-5}$ & $7.4\pm0.5$ & $144\pm23$ &  $0.04013\pm0.0003$\\
		$0.56497\pm2.0\times10^{-5}$ & $1.7473\pm1.0\times10^{-4}$ & $6.4\pm0.5$ & $172\pm34$ &  $0.03991\pm0.0002$\\
		$0.57265\pm1.6\times10^{-5}$ & $1.7792\pm9.0\times10^{-5}$ & $5.1\pm0.4$ & $230\pm46$ &  $0.03964\pm0.0002$\\
		$0.57995\pm1.5\times10^{-5}$ & $1.8122\pm9.0\times10^{-5}$ & $4.0\pm0.3$ & $272\pm61$ &  $0.03952\pm0.0001$\\
		$-0.52443\pm1.6\times10^{-5}$ & $1.5837\pm7.0\times10^{-5}$ & $8.1\pm0.3$ & $137\pm12$ &  $-0.04009\pm0.0002$\\
		$-0.53174\pm1.4\times10^{-5}$ & $1.6127\pm7.0\times10^{-5}$ & $7.2\pm0.4$ & $153\pm20$ &  $-0.03990\pm0.0002$\\
		$-0.53787\pm1.6\times10^{-5}$ & $1.6382\pm9.0\times10^{-5}$ & $6.1\pm0.4$ & $188\pm29$ &  $-0.03971\pm0.0002$\\
		$-0.54429\pm1.6\times10^{-5}$ & $1.6653\pm1.0\times10^{-4}$ & $5.0\pm0.4$ & $227\pm39$ &  $-0.03956\pm0.0001$\\
		$-0.55005\pm1.6\times10^{-5}$ & $1.6915\pm9.0\times10^{-5}$ & $3.7\pm0.4$ & $308\pm65$ &  $-0.03944\pm0.0001$\\
		\hline
		$0.40583\pm1.8\times10^{-5}$ & $0.9111\pm8.8\times10^{-5}$ & $9.8\pm0.7$ & $211\pm39$ &  $0.02916\pm0.0002$\\
		$0.40978\pm2.7\times10^{-5}$ & $0.9234\pm1.1\times10^{-4}$ & $9.1\pm0.3$ & $198\pm14$ &  $0.02905\pm0.0001$\\	
		$0.41341\pm2.4\times10^{-5}$ & $0.9334\pm8.8\times10^{-5}$ & $8.0\pm0.2$ & $233\pm18$ & $0.02888\pm0.00004$\\			
		$0.41621\pm1.7\times10^{-5}$ & $0.9402\pm7.0\times10^{-5}$ & $6.77\pm0.09$ & $272\pm4$ &  $0.02870\pm0.00003$\\	
		$0.41898\pm2.1\times10^{-5}$ & $0.9483\pm8.5\times10^{-5}$ & $5.65\pm0.09$ & $330\pm15$ &  $0.02857\pm0.00003$\\	
		$-0.39226\pm2.2\times10^{-5}$ & $0.8714\pm1.0\times10^{-4}$ & $9.7\pm0.5$ & $210\pm26$ &  $-0.02909\pm0.00013$\\
		$-0.39558\pm2.2\times10^{-5}$ & $0.8810\pm8.2\times10^{-5}$ & $8.7\pm0.3$ & $225\pm20$ &  $-0.02894\pm0.00006$\\	
		$-0.39828\pm1.9\times10^{-5}$ & $0.8884\pm6.8\times10^{-5}$ & $7.6\pm0.2$ & $258\pm13$  & $-0.02879\pm0.00003$\\
		$-0.40065\pm1.6\times10^{-5}$ & $0.8944\pm6.1\times10^{-5}$ & $6.38\pm0.03$ & $307\pm5$ & $-0.02864\pm0.00003$\\	
		$-0.40238\pm1.9\times10^{-5}$ & $0.89901\pm8.1\times10^{-5}$ & $5.4\pm0.2$ & $354\pm26$ &  $-0.02854\pm0.00003$\\																
		\hline\hline
	\end{tabular}	
	\caption{ Parameters of the solitary wave extracted from micromagnetic simulations. Superscript $^{\rm m}$ marks the solitary wave whose dynamics is illustrated in the movie {\fontfamily{qcr}\selectfont movie2.mkv}, see Fig.~\ref{fig:movie}.}\label{tbl:raw-data}
\end{table*}

\bibliography{string}